\documentclass[a4paper, twoside, reqno]{amsart}

\usepackage{amsthm}
\usepackage{amssymb}
\usepackage[foot]{amsaddr}
\usepackage{graphicx}
\usepackage{marvosym}
\usepackage{textcase}
\usepackage[
    backend=biber,
    natbib=true,
    citestyle=authoryear-comp,
    bibstyle=authoryear,
    maxcitenames=2,
    maxbibnames=9,
    uniquelist=false,
    uniquename=false,
    dashed=false,
    firstinits=true,
    doi=false,
    isbn=false,
    date=year]{biblatex}
\usepackage[
    hidelinks,
    colorlinks=true,
    allcolors=blue,
    pdfproducer={Latex with hyperref},
    pdfcreator={pdflatex}]{hyperref}
\usepackage{bm}
\usepackage{glossaries}

\newcommand{\papertitle}{Rejoinder to the discussion of ``Mode-based estimation of the center of symmetry''}
\newcommand{\runningtitle}{Rejoinder}
\newcommand{\metadatatitle}{Rejoinder to the discussion of "Mode-based estimation of the center of symmetry"}

\newcommand{\myorcid}[1]{%
    \href{https://orcid.org/#1}{%
        \includegraphics[height=2ex]{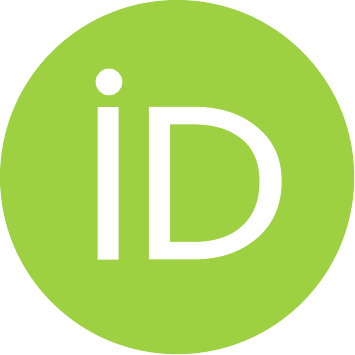}%
        \url{https://orcid.org/#1}
    }%
}

\newcommand{\joseechacon}{Jos\'e E. Chac\'on}

\newcommand{\addressjoseechaon}{Departamento de Matem\'aticas, Universidad de Extremadura, Badajoz, Spain.}

\newcommand{\emailjoseechacon}{jechacon@unex.es}
\newcommand{\orcidjoseechacon}{0000-0002-3675-1960}
\newcommand{\footjoseechacon}{$^{\dagger}$}
\newcommand{\footaddressjoseechaon}{\footjoseechacon\addressjoseechaon}
\newcommand{\footorcidjoseechacon}{\footjoseechacon\myorcid{\orcidjoseechacon}}

\newcommand{\javierfdezserrano}{Javier Fern\'andez Serrano}

\newcommand{\addressjavierfdezserrano}{Departamento de Matem\'aticas, Universidad Aut\'onoma de Madrid, Madrid, Spain.}

\newcommand{\emailjavierfdezserrano}{javier.fernandezs01@estudiante.uam.es}
\newcommand{\orcidjavierfdezserrano}{0000-0001-5270-9941}
\newcommand{\footjavierfdezserrano}{$^{\ddagger}$}
\newcommand{\footaddressjavierfdezserrano}{\footjavierfdezserrano\addressjavierfdezserrano}
\newcommand{\footorcidjavierfdezserrano}{\footjavierfdezserrano\myorcid{\orcidjavierfdezserrano}}

\newcommand{\hino}{Hino}
\newcommand{\pardofernandez}{Pardo-Fernández}

\newcommand{\mscnonparamestimation}{62G05}
\newcommand{\mscdensityestimation}{62G07}
\newcommand{\mscnonparamrobustness}{62G35}

\newcommand{\goesto}{\rightarrow}
\newcommand{\vectordimension}{d}
\newcommand{\reals}{\mathbb{R}}
\newcommand{\rd}{\reals^{\vectordimension}}
\newcommand{\weights}{W_i}
\newcommand{\kernel}{K}
\newcommand{\bandwidth}{h}
\newcommand{\shapeparameter}{\beta}
\newcommand{\chaconserranokernelof}[1]{\kernel_{#1}}
\newcommand{\chaconserranokernel}{\chaconserranokernelof{\shapeparameter}}
\newcommand{\trimminglevel}{\alpha}

\input{configuration.tex}
\newacronym[
    plural=KMEs,
    firstplural=kernel mode estimators]{kme}{KME}
{kernel mode estimator}

\begin{document}

\title[\runningtitle]{\papertitle}
\date{}

\hypersetup{
  pdftitle={\metadatatitle},
}

\author{\joseechacon\footjoseechacon}
\address{\footaddressjoseechaon}
\email{\footjoseechacon\emailjoseechacon \ \Letter}
\thanks{\footorcidjoseechacon}

\author{\javierfdezserrano\footjavierfdezserrano}
\address{\footaddressjavierfdezserrano}
\email{\footjavierfdezserrano\emailjavierfdezserrano}
\thanks{\footorcidjavierfdezserrano}

\hypersetup{
    pdfauthor={\joseechacon, \javierfdezserrano}
}

\begin{abstract}
    Rejoinder to the discussion by \citet{Hino2025} and \citet{PardoFernandez2025} of \citet{Chacon2025}, a special paper (with discussion) published in \textit{Annals of the Institute of Statistical Mathematics}.
\end{abstract}

\hypersetup {
    pdfsubject={%
            Rejoinder to the discussion by Hino (2025) and Pardo-Fernández (2025) of Chacón and Fernández Serrano (2025), a special paper (with discussion) published in Annals of the Institute of Statistical Mathematics.%
        }%
}

\subjclass[2020]{
    \mscnonparamestimation \ (Primary),
    \mscdensityestimation,
    \mscnonparamrobustness
}

\maketitle

We want to express our sincere gratitude to Professor \hino{} and Professor \pardofernandez{} for their kind words and for providing such insightful and stimulating discussions.
We are also extremely honored to have the opportunity to expand on some of their comments in this rejoinder.
Hopefully, this series of papers will foster new developments in statistics from a modal perspective.

\section{Professor \hino{}'s comments}

Professor \hino{} provides an excellent summary of our core contributions.
His discussion rightly emphasizes the connection between the \gls*{kme} and robust M-estimators, as well as the novel finding that both kernel shape and bandwidth significantly impact the performance of the \gls*{kme} in the symmetric, unimodal setting.

In addition, Professor \hino{} conveniently points out alternative mode estimation techniques, such as $k$-NN-based methods and the half-sample mode.
These are indeed important tools in the broader landscape of mode estimation.
However, our paper focuses on kernel-based methods because they enjoy a unique property with symmetric data: producing unbiased estimates.
The latter allows employing non-vanishing bandwidths designed to minimize the asymptotic variance.
In this respect, it would be interesting to investigate if other mode estimators behave similarly.

On the other hand, we share Professor \hino{}'s enthusiasm about the wide range of applications of the mode \citep{Chacon2020}.
His works on modal linear regression \citep{Sando2019} and modal principal component analysis \citep{Sando2020} are great examples of how the mode can secure robustness in classic problems.

\section{Professor \pardofernandez{}'s comments}

Professor \pardofernandez{} raises three specific and relevant potential extensions of our work, dealing with censored/truncated data, multivariate settings, and testing for symmetry.
We briefly comment on each of them in the following.

\subsection{Censoring and truncation}

Censoring and truncation are pertinent subjects related to our research.
As Professor \pardofernandez{} notes, estimators in these settings, like the Kaplan-Meier or Lynden-Bell estimators, involve random weights $\weights$ that depend on the full sample, unlike the uniform $1/n$ weights in the standard i.i.d. case.
Extending our \gls*{kme} framework would require adapting the kernel density estimator definition (as in Equation (2) from the discussion) using these weighted distribution function estimators.

Certainly, \glspl*{kme} have been studied under censoring and truncation \citep[see][and references therein]{Guessoum2018}.
However, to our knowledge, the consequences of the additional symmetry assumption in these settings have \textit{not} been investigated so far.
In that context, the primary challenges would lie, on one hand, in deriving the asymptotic variance of the \gls*{kme}, and, on the other hand, in designing a suitable version of the iterative reweighting algorithm.
For the former, the random weights $\weights$ would imply a considerably more complex scenario due to their intricate dependence structure; for the latter, the additional layer of $\weights$s might affect the stability, convergence properties, and interpretation of the algorithm.

Nevertheless, this is a valuable direction, as robust estimation under such data limitations is crucial in many fields (e.g., survival analysis or econometrics).
Accordingly, this problem needs---and deserves---to be examined in greater depth in future publications.

\subsection{The multivariate case}

As Professor \pardofernandez{} suggests, extending the theory to the multivariate setting is a natural and important next step.
Nonetheless, this problem appears challenging since our results rely on two essential concepts, symmetry and unimodality, for which several possible generalizations exist.

A multivariate density function $f\colon\rd\goesto\reals$ can be symmetric in many ways.
Considering the origin as the center of symmetry, numerous definitions exist, including central symmetry, axial symmetry, rotational symmetry, radial symmetry, or elliptical symmetry, to name only a few.
All these notions can be shifted to an arbitrary center of symmetry $\bm{\theta} \in \rd$ through a simple translation.
See \citet{Serfling2006} for an interesting survey on multivariate symmetry.

Regarding unimodality, \citet{Dharmadhikari1988} study various non\-equivalent approaches in the multivariate context: star unimodality, block unimodality, linear unimodality or (central) convex unimodality, among others.
They also show that the class of log-concave densities, which has recently been the focus of major research interest \citep[for a review, see][]{Samworth2018}, can play an important role within these unimodal distributions.

Therefore, developing the corresponding multivariate theory necessarily includes, as a first task, identifying the appropriate class of symmetric and unimodal densities for which the convolution with a kernel leaves the mode invariant.
That is the key feature that allows, in a second stage, focusing on deriving the asymptotic variance of the \gls*{kme}.
Finally, obtaining a flexible parametric family of multivariate kernels (akin to our $\chaconserranokernel$), to fully leverage the efficiency and robustness of this nonparametric approach, surely represents a nontrivial task.

\subsection{Symmetry testing}

Using our \gls*{kme} to enhance the capabilities of existing symmetry tests is a very nice suggestion by Professor \pardofernandez{}.
Since our optimized \gls*{kme} can offer higher efficiency in estimating the center of symmetry, incorporating it into a symmetry test statistic could lead to a significant power increase in some cases.

The trimmed mean was selected by \citet{Milosevic2018} because of the flexibility of its trimming level $\trimminglevel \in (0, 1/2)$, having the sample mean and the sample median as limiting cases as $\trimminglevel \goesto 0$ and $\trimminglevel \goesto 1/2$, respectively.
Similarly, in vanilla \gls*{kme}, the role of $\trimminglevel$ is played by the bandwidth $\bandwidth$.
Furthermore, the resemblance between the \gls*{kme} and the trimmed mean when using the Epanechnikov kernel was noted in our work.

However, more importantly, the \gls*{kme} based on the parametric family $\chaconserranokernel$ supplies additional flexibility with the shape parameter.
Taking $\shapeparameter \goesto \infty$ allows retrieving the Epanechnikov kernel, whereas making $\shapeparameter \goesto 0$ produces median-like behavior.
Moreover, interestingly, our variance-minimizing process favored small intermediate values $\shapeparameter < 1$ under heavier tails such as those of the Cauchy and the logistic \citep[two of the three test-beds proposed in the study by][]{Milosevic2018}.
In such scenarios, symmetry testing based on our full \gls*{kme} proposal could be more efficient than based on the trimmed mean.

\section*{Acknowledgements}

The research of the first author has been supported by the MICINN grant PID2021-124051NB-I00, while both authors have been supported by the MICINN grant PID2023-148081NB-I00.
We want to thank everyone involved in publishing this series of articles.
Special thanks to the two discussants, Professor \hino{} and Professor \pardofernandez{}, and to the chief editor, Professor Ninomiya, for his decision to turn our original manuscript into a special paper.

\renewcommand*{\mkbibcompletename}[1]{\textsc{#1}}
\printbibliography[filter=references, title={References}, sorting=nyt]

\end{document}